\documentclass{article}
\usepackage{spconf}
\usepackage{amsmath,amssymb,amsfonts,cite,epsfig, epstopdf}

\usepackage{graphicx}
\usepackage{subfigure}
\usepackage{color}
\usepackage{verbatim}

\title{A CFAR Adaptive Matched Detector for Target Detection in Non-Gaussian Noise With Inverse Gamma Texture}

\name{Shiwen~Lei$^*$, Andreas~Jakobsson$^*$, and Zhiqin~Zhao$^\dagger$
\thanks{This work was supported in part by the Swedish Research Council and Carl Trygger's foundation.}}
\address{
$^*$Dept. of Mathematical Statistics, Lund University, Sweden\\
$^\dagger$School of Electronic Engineering, University of Electronic Science and Technology of China, China
}

\begin{document}
\maketitle

\begin{abstract}
In this paper, we propose an adaptive matched  detector of a signal corrupted by a non-Gaussian noise with an inverse gamma texture. The detector is formed using a set of secondary data measurements, and is analytically shown to have a constant false alarm rate.
The analytic performance is validated using Monte Carlo simulations, and the proposed detector is shown to offer preferable performance as compared to the related
one-step generalized likelihood ratio test (\mbox{1S-GLRT}) and the adaptive subspace detector (ASD).
%
%
\end{abstract}

\section{Introduction}

The problem of detecting a partly known target corrupted by an additive noise is commonly occurring in a variety of fields, such as, for instance, radar and sonar applications.
Early works focused on the case of homogeneous noise fields, where in the noise in in different test cells was assumed to have the same statistical properties, introducing classical detectors such as the generalized likelihood ratio tests (GLRTs) \cite{Kelly86_22, Park95_41} and the adaptive matched filters (AMFs) \cite{Robey92_28, DeMaio01_81}.
These works were later extended to cases wherein the primary and secondary data are allowed to gave different statistical properties, and detectors such as the
matched subspace detectors (MSDs) \cite{ScharfF94_42, Burgess96_44} and the adaptive subspace detectors (ASDs)\cite{KrautSM01_49, LiuZYL11_59} were introduced.
Of these, the former assumes that the noise covariance matrix (NCM) is known {\em a priori}, whereas the latter estimates the NCM using secondary data.
%
%
In cases when the background noise can no longer be assumed homogeneous, or even partially homogeneous, such as in target detection in a sea or earth background, one often the noise to be non-Gaussian, using an inverse gamma texture model \cite{Akcakaya11_47, Javier11_47, Sangston12_48}. In such cases, the noise is typically assumed to be formed by two independent parameters, namely that of the texture, $\kappa$, and the speckle, $\mathbf{g}$ \cite{WangLH13_61, LiCKY12_6, Hurtado08_56}.
Well known contributions to this problem includes the texture-free GLRT (TF-GLRT) \cite{Pastina01_37}, which does not consider the influence of $\kappa$, and the one-step GLRT (\mbox{1S-GLRT}), the two-step GLRT (2S-GLRT), and the maximum $\textit{a posteriori}$ GLRT (MAP-GLRT) detectors \cite{ShangS11_5}.
The three latter have the same test statistic, but their exact performance are complicated to calculate. In cases when the statistical properties in different channels are the same, the \mbox{TF-GLRT} coincides with the \mbox{1S-GLRT}.
Further extensions include detectors taking into account the persymmetric property of the NCM, offering improved performance in case of non-Gaussian noise \cite{GaoLZY13_20, Pailloux11_47,  DeMaioO15_63}. However, this improvement strongly relies on the symmetric distribution of the measurement array.
%
%
In this work, we strive to include the the influence of $\kappa$, designing a matched detector in the case of non-Gaussian noise, without imposing the persymmetric assumption. We derive the exact performance probabilities for both deterministic and fluctuating targets, showing that the proposed estimator has a constant false alarm rate (CFAR). The accuracy of the presented probabilities are verified using numerical simulations, and the effectiveness of the proposed detector is assessed by comparing with the \mbox{1S-GLRT} \cite{ShangS11_5} and the ASD \cite{LiuZYL11_59}.


%
\begin{figure}[t!]
\centering
	\subfigure[]{\includegraphics[width=0.8\columnwidth]{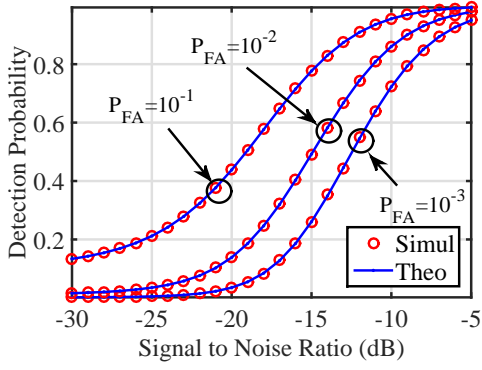}} \vspace{-7mm}�\\
	\subfigure[]{\includegraphics[width=0.8\columnwidth]{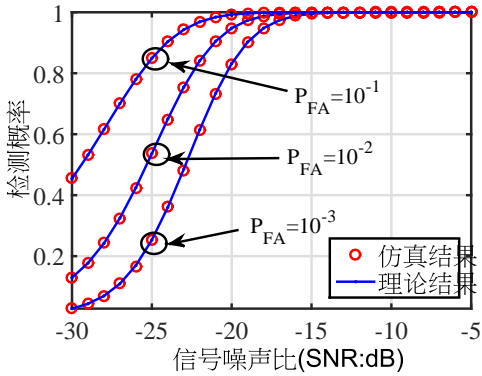}} \vspace{-4mm}
\caption{ Detection probability for deterministic target detection 
with: (a) ($\alpha,\beta$)=(2,0.5) and (b) ($\alpha,\beta$)=(5,2).}\vspace{-1mm}
\label{Fig. 1}
\end{figure}
\begin{figure}[t!]
\centering
	\subfigure[]{\includegraphics[width=0.8\columnwidth]{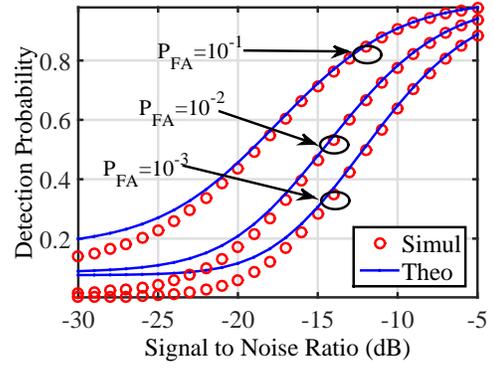}} \vspace{-7mm} \\
	\subfigure[]{\includegraphics[width=0.8\columnwidth]{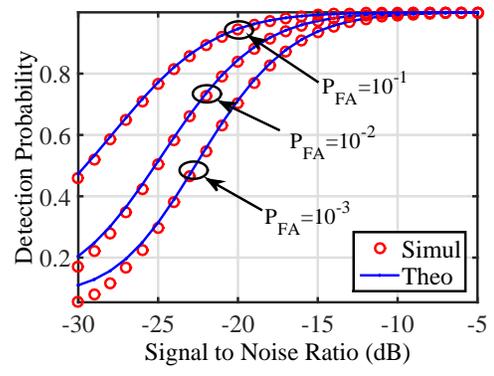}} \vspace{-4mm}
\caption{ Detection probability for fluctuating target detection 
with: (a) ($\alpha,\beta$)=(2,0.5) and (b) ($\alpha,\beta$)=(5,2).}\vspace{-1mm}
\label{Fig. 2}
\end{figure}
%

\section{Adaptive Matched  Detection in Non-Gaussian noise}

Consider the $N\times1$  measurement vector
$\mathbf{y}=\mathbf{Ax}+\mathbf{v}$,
where
$\mathbf{A}$ and $\mathbf{x}$ denote the known $N\times r$ dimensional system response matrix, with $r \ll N$, and the $r\times1$ dimensional target echo, respectively, and with $\mathbf{v}$ denoting a $N\times1$ dimensional additive non-Gaussian noise.
The noise is assumed to have an inverse gamma texture, such that
$\mathbf{v}=\sqrt{\kappa}\mathbf{g}$,
where the texture parameter, $\kappa > 0$, and the speckle parameter, $\mathbf{g}$, are assumed independent. The texture is assumed to follow an inverse Gamma distribution with shape  parameter $\alpha$ and scaling parameter $\beta$, having the PDF 
\begin{equation}\label{eq:f_IG}
f_{IG}(\kappa; \alpha, \beta)=\dfrac{1}{\beta^{\alpha}\Gamma(\alpha)}\kappa^{-(\alpha+1)}\exp\left(-\dfrac{1}{\beta\kappa}\right) 
\end{equation}
where $\Gamma(\alpha)=\int_{0}^{+\infty}u^{\alpha-1}e^{u}du$ denotes the Gamma function.
Furthermore, the speckle, describing the structural information, is assumed to be zero mean and have the same NCM in adjacent cells, i.e., $\mathbf{g}\sim CN(0,\textbf{R})$.
The detection problem of interest may thus be formulated as  the binary hypothesis test
\begin{align}\label{eq:H0H1}
\begin{cases}
H_0: \mathbf{y}=\mathbf{v}\sim CN(0,\kappa\mathbf{R})\\
H_1: \mathbf{y}=\mathbf{Ax}+\mathbf{v}\sim CN(\mathbf{Ax},\kappa\mathbf{R})
\end{cases}
\end{align}
where the NCM is typically formed using $K$ secondary data vectors, using, for instance, the normalized sample covariance matrix (NSCM), i.e.,
\begin{align}\label{eq:Rhat}
\hat{\mathbf{R}} = \dfrac{N}{K}\sum_{k=1}^K\dfrac{\mathbf{y}_k\mathbf{y}_k^H}{\mathbf{y}_k^H\mathbf{y}_k}
\end{align}
where $\mathbf{y}_k$ denotes the $k$:{th} secondary data, and $(\cdot)^H$ the Hermitian conjugate transpose.
From \eqref{eq:H0H1}, the PDFs of the measurement are
\begin{align}
\begin{cases}
H_0: \textit{f}_0(\mathbf{y})=\dfrac{1}{\pi^{N}\kappa^N|\mathbf{R}|}\text{etr}(\kappa^{-1}\mathbf{R}^{-1}\textbf{T}_0)\\
H_1: \textit{f}_1(\mathbf{y})=\dfrac{1}{\pi^{N}\kappa^N|\mathbf{R}|}\text{etr}(\kappa^{-1}\mathbf{R}^{-1}\textbf{T}_1)
\end{cases}
\end{align}
where $\text{etr}(\mathbf{A}) = \exp( \mbox{tr}\!\left\{\mathbf{A}\right\})$,
%
$\mathbf{T}_0 =\mathbf{y}\mathbf{y}^H$, and
$\mathbf{T}_1 =(\mathbf{y}-\mathbf{Ax})(\mathbf{y}-\mathbf{Ax})^H$.
The test statistic may thus be formed as
%
\begin{equation}\label{glrt}
\Lambda=\dfrac{\max_{\mathbf{x}} f_1(\mathbf{y})}{f_0(\mathbf{y})}\overset{\textit{H}_1}{\underset{\textit{H}_0}{\gtrless}}\Lambda_0
\end{equation}
with
$\Lambda_0$ denoting the detection threshold.
Setting the first-order derivation of $f_1(\mathbf{y})$ with respect to (w.r.t.) $\mathbf{x}$ equal to zero, the MLE of $\mathbf{x}$ may be formed as 
\begin{equation}
\hat{\mathbf{x}}=(\mathbf{A}^H\mathbf{R}^{-1}\mathbf{A})^{-1}\mathbf{A}^H\mathbf{R}^{-1}\mathbf{y}
\end{equation}
which, if substituted into \eqref{glrt}, and replacing $\mathbf{R}$ with $\hat{\mathbf{R}}$, as given by \eqref{eq:Rhat}, yields the non-Gaussian adaptive matched detector (\mbox{nG-AMD}) as
%
%
\begin{equation}\label{eq:Lambda}
\Lambda=\mathbf{y}^H\hat{\mathbf{R}}^{-1}\mathbf{A}(\mathbf{A}^H\hat{\mathbf{R}}^{-1}\mathbf{A})^{-1}\mathbf{A}^H\hat{\mathbf{R}}^{-1}\mathbf{y}\overset{\textit{H}_1}{\underset{\textit{H}_0}{\gtrless}}\Lambda_0
\end{equation}
%
It is worth noting that the \mbox{nG-AMD} has the same form as the AMD for detecting target in partially homogeneous background \cite{LeiZNL15_104}, although the latter assumes a constant texture, whereas
\mbox{nG-AMD} allows for the texture to vary.

\begin{figure}[t!]
\centering
	\subfigure[]{\includegraphics[width=0.8\columnwidth]{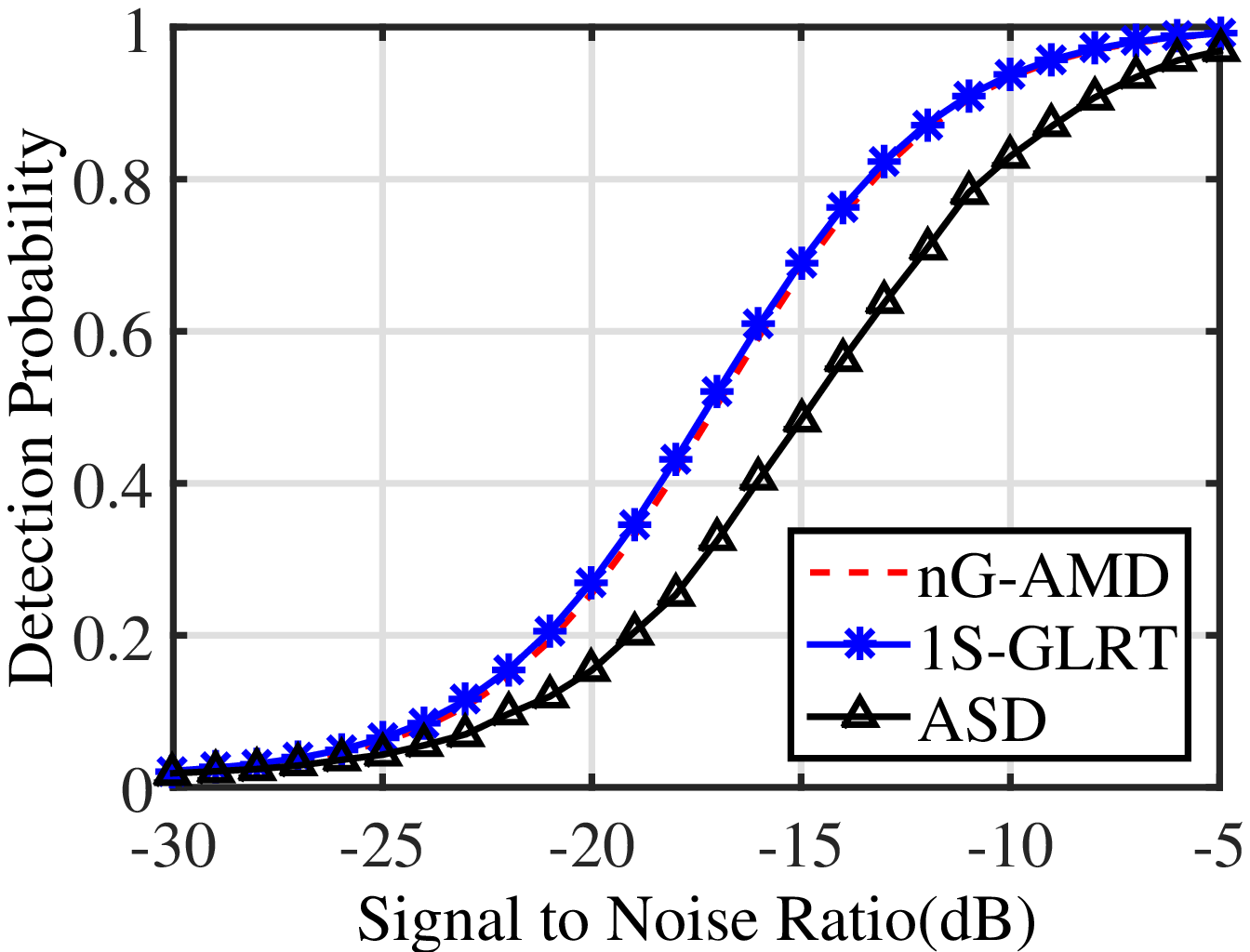}} \vspace{-8mm} \\
	\subfigure[]{\includegraphics[width=0.8\columnwidth]{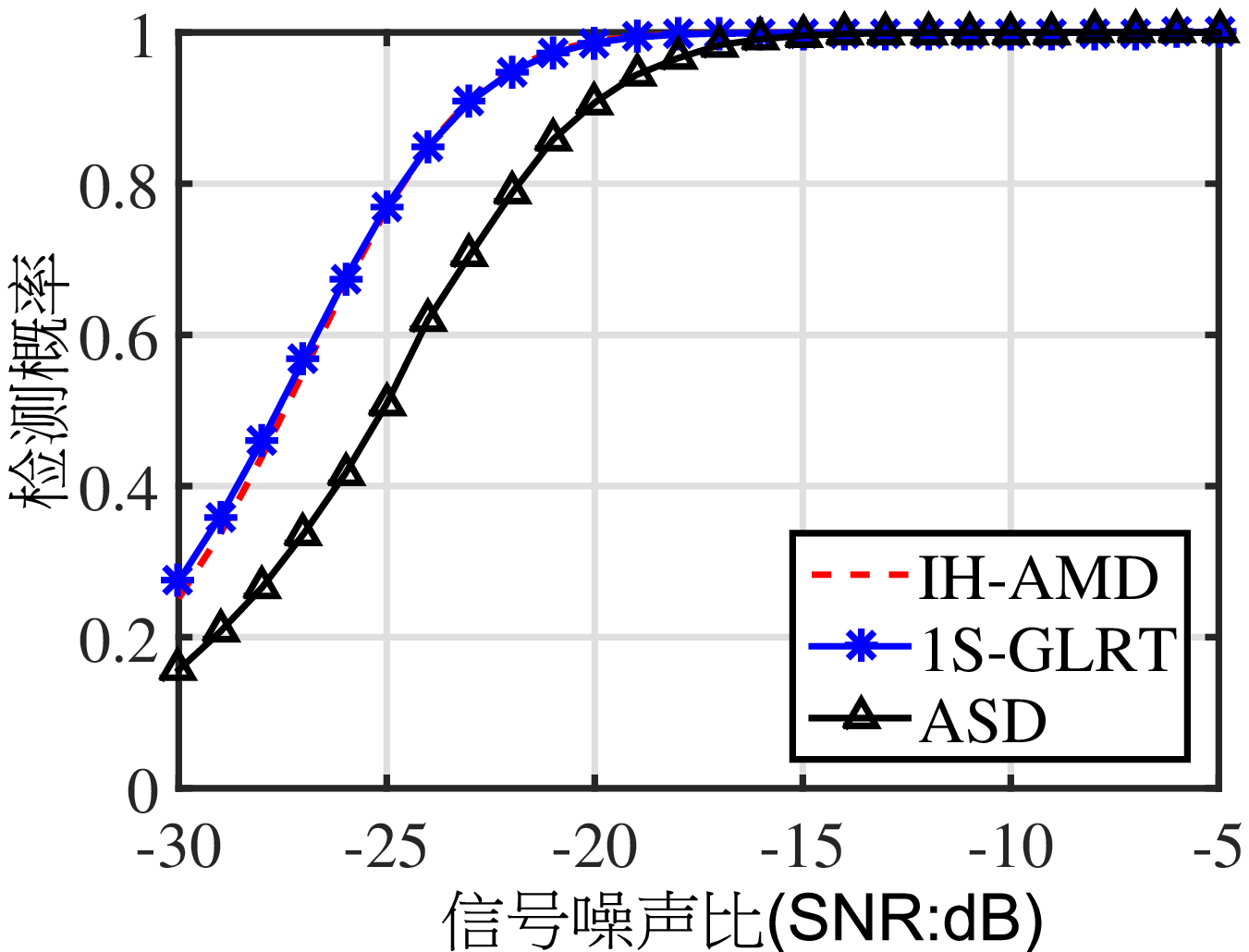}} \vspace{-5mm}
\caption{Detection probabilities for a deterministic target, with $P_\text{FA}=10^{-2}$: (a) ($\alpha,\beta$)=(2,0.5) and (b) ($\alpha,\beta$)=(5,2).}\vspace{-1mm}
\label{Fig. 3}
\end{figure}%

\section{Theoretical Detection Performance}

Since both deterministic target and fluctuating target commonly appear in target detection problems (see, e.g., \cite{LiuZYL11_59, JinF05_53, Jayaweera07_55}), we will here consider the detection performance of \mbox{nG-AMD} in both these cases.
As shown in 
[20], the considered test statistic will follow an $F$ distribution, and may thus be expressed as the quotient of two independent chi-square distributions, such that
\begin{equation}\label{eq:Fdist}
\rho\Lambda=
\begin{cases}
\dfrac{\chi^2_{2r}}{\chi^2_{2(K-N+1)}},~~~~~\text{under}~~\textit{H}_0\\
\dfrac{\chi^2_{2r}(\mu_0)}{\chi^2_{2(K-N+1)}},~~~~~\text{under}~~\textit{H}_1
\end{cases}
\end{equation}
where $\mu_0=2\rho\mathbf{x}^H\mathbf{A}^H(\kappa\mathbf{R})^{-1}\mathbf{Ax}$, and $\chi^2_n$ represents a chi-square distribution function with $n$ degrees of freedom with non-central parameter $\mu_0$, with $\mu_0 = 0$ if not specified.
%
%
Let $2t_\ell$ and $2\tau$ denote the numerator, under hypothesis $\ell$, and the dominator in \eqref{eq:Fdist}, respectively.
%
Then, the PDFs of $t_\ell$ and $\tau$, i.e., $f_t(t|\textit{H}_0)$, $f_t(t|\textit{H}_1)$, and $f_\tau(\tau)$ will be given by (A.23), (A.24), and (A.25) in [20], respectively. Similarly, the PDF of $\rho$, i.e., $f_\rho(\rho)$, will be given by (16) in [20].
The test statistic in \eqref{eq:Lambda} may thus be rewritten as
$t{\gtrless}\tau\rho\Lambda_0$.
%
%
Let
\begin{equation}\label{eq:mu}
\mu_1=\mathbf{x}^H\mathbf{A}^H\hat{\mathbf{R}}^{-1}\mathbf{Ax},~~~~~\mu=\mu_0/(2\rho\kappa)=\mu_1/\kappa
\end{equation}
Then, using \eqref{eq:f_IG}, the PDF of  $\mu$ is given as
\begin{equation}
f_{\mu}(\mu)=\dfrac{1}{(\beta\mu_1)^\alpha\Gamma(\alpha)}\mu^{\alpha-1}\exp\left(-\dfrac{\mu}{\beta\mu_1}\right)
\end{equation}
where $\mu>0$.
According to \eqref{eq:Fdist}, the false alarm probability of \mbox{nG-AMD}, here denoted $P_\text{FA}^\text{\mbox{nG-AMD}}$, will thus depends on the system dimension, $N$, and the signal dimension, $r$, but not on the noise, and therefore has a CFAR.
The false alarm probability may be calculated as
%
\begin{equation}\label{eq:pFA}
P_\text{FA}^\text{\mbox{nG-AMD}}=\int_{0}^{1}P_{\text{FA}|\rho}^\text{\mbox{nG-AMD}}f_\rho(\rho)d\rho
\end{equation}F
where the conditional false alarm probability, $P_{\text{FA}|\rho}^\text{\mbox{nG-AMD}}$, is
\begin{align*}
P_{\text{FA}|\rho}^\text{\mbox{nG-AMD}}&=\int_{0}^{+\infty}\int_{\tau\rho\Lambda_0}^{+\infty}f_t(t|\textit{H}_0)dtf_{\tau}(\tau|\textit{H}_0)d\tau\notag\\
& \hspace{-9mm} =\dfrac{1}{ (1+\rho\Lambda_0)^{K-N+1} }\sum_{i=1}^{r}C_{K-N+r-i}^{r-i} \dfrac{ (\rho\Lambda_0)^{r-i}}{(1+\rho\Lambda_0)^{r-i} }
\end{align*}
with $C_n^m=n!/(m!(n-m)!)$ denoting the binomial coefficients.
%
%
We proceed to determine the probability of detection for a deterministic target, i.e., for the case when $\mu_1$, as defined in \eqref{eq:mu}, is deterministic. Under $\textit{H}_1$, the conditional detection probability $P_{D|\mu,\rho}^\text{\mbox{nG-AMD}}$ is then
\begin{align*}
P_{D|\mu,\rho}^\text{\mbox{nG-AMD}}&=\int_{0}^{+\infty}\int_{\omega}^{+\infty}f_t(t|\textit{H}_1)dtf_{\omega|\rho}(\omega)d\omega\notag\\
& \hspace{-9mm} = 1 - \dfrac{ (\rho\Lambda_0)^r }{ (1+\rho\Lambda_0 )^{r+K-N} }
\sum_{i=0}^{K-N}C_{K-N+r}^{r+i}\notag\\
& \hspace{-9mm} ~~~\times(\rho\Lambda_0)^i\exp\left(-\dfrac{\mu\rho}{1+\rho\Lambda_0}\right)\sum_{m=0}^i\dfrac{1}{m!}\left(\dfrac{\mu\rho}{1+\rho\Lambda_0}\right)^m
\end{align*}
where $\omega=\tau\rho\Lambda_0$, and
\begin{align}
f_{\omega|\rho}(\omega)
=\dfrac{1}{(k-N)!}\dfrac{1}{\rho\Lambda_0}(\dfrac{\omega}{\rho\Lambda_0})^{K-N}\exp{ \left(-\dfrac{\omega}{\rho\Lambda_0} \right) }
\end{align}
Hence, the conditional detection probability, $P_{D|\rho}^{\textit{\mbox{nG-AMD}}}$, may be expressed as
\begin{align}\label{PDrho}
P_{D|\rho}^\text{\mbox{nG-AMD}}&=\int_{0}^{+\infty}P_{D|\mu,\rho}^\text{\mbox{nG-AMD}}f_\mu(\mu)d\mu\notag\\
&  \hspace{-9mm} =1-\dfrac{1}{(\beta\Lambda_0)^\alpha\Gamma(\alpha)}(\dfrac{\rho\Lambda_0}{1+\rho\Lambda_0})^r(\dfrac{1}{1+\rho\Lambda_0})^{K-N}\notag\\
&  \hspace{-9mm} ~~\times\sum_{i=0}^{K-N}C_{K-N+r}^{r+i}(\rho\Lambda_0)^i\sum_{m=0}^i\dfrac{1}{m!}(\dfrac{\rho}{1+\rho\Lambda_0})^m\notag\\
&  \hspace{-9mm} ~~\times(\dfrac{\rho}{1+\rho\Lambda_0}+\dfrac{1}{\beta\mu_1})^{-(m+\alpha)}\Gamma(m+\alpha)
\end{align}
yielding the detection probability of \mbox{nG-AMD} for deterministic target detection
%
\begin{figure}[t!]
\centering
	\subfigure[]{\includegraphics[width=0.8\columnwidth]{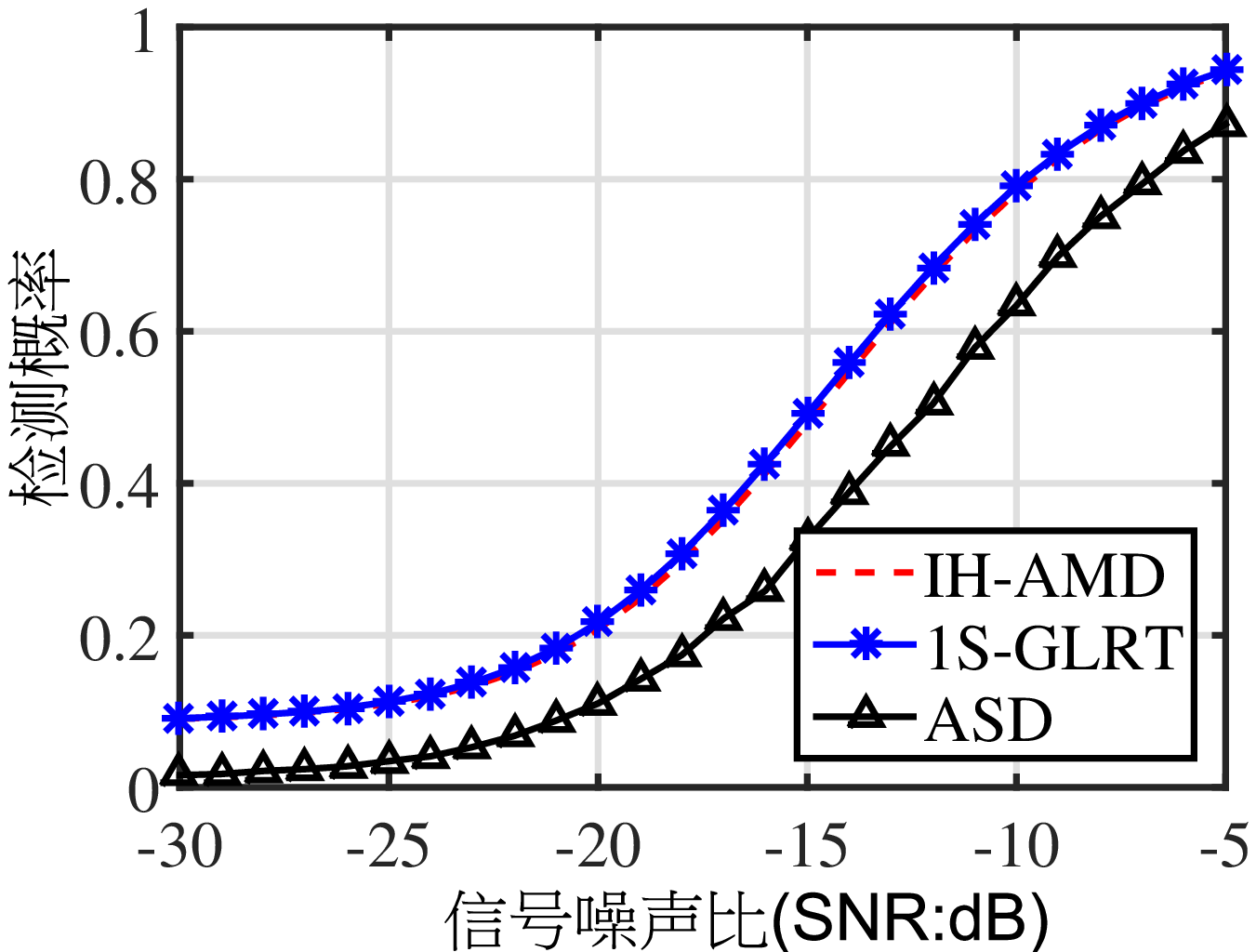}} \vspace{-8mm} \\
	\subfigure[]{\includegraphics[width=0.8\columnwidth]{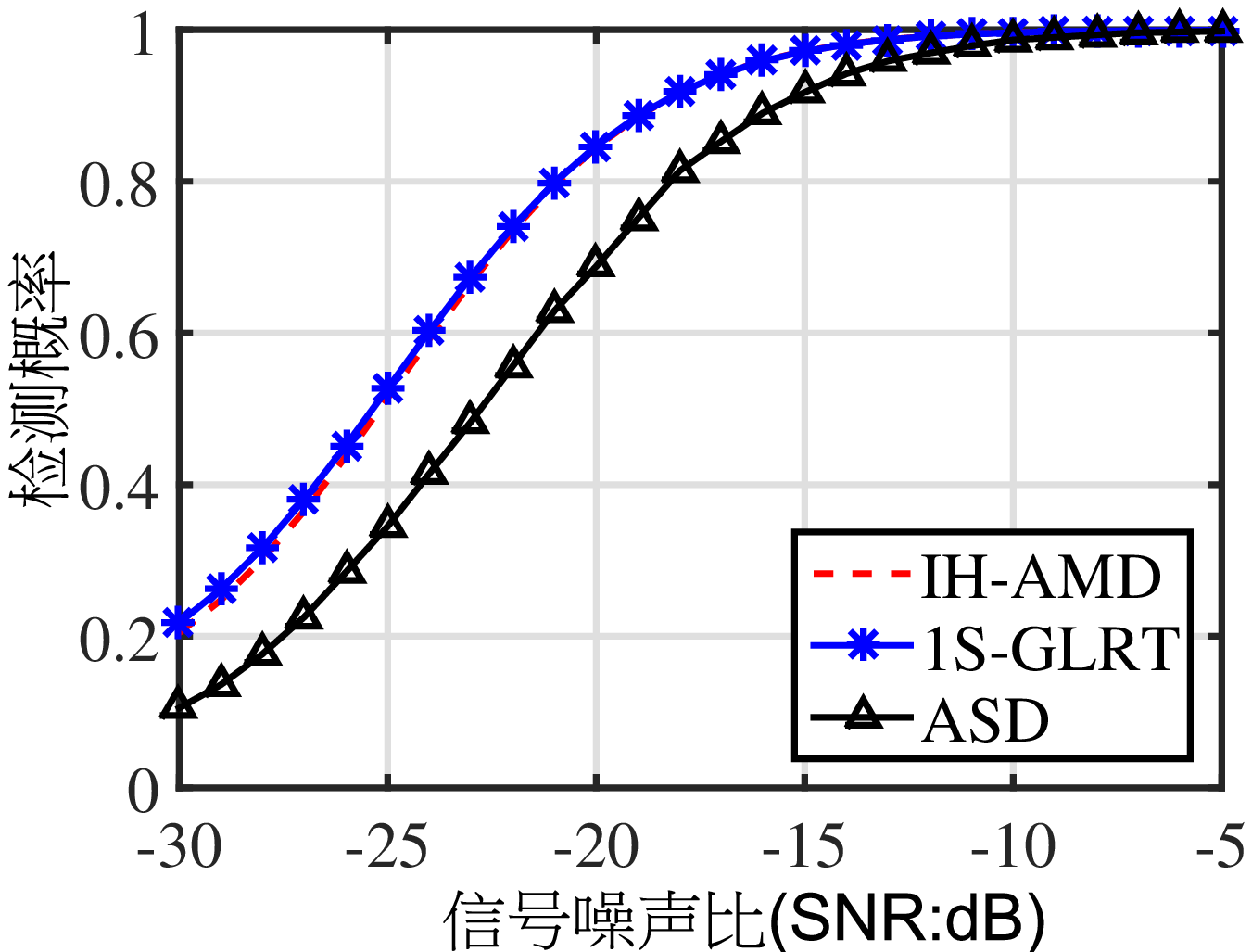}} \vspace{-5mm}
\caption{Detection probabilities for a fluctuating target, with $P_\text{FA}=10^{-2}$: (a) ($\alpha,\beta$)=(2,0.5) and (b) ($\alpha,\beta$)=(5,2).}\vspace{-3mm}
\label{Fig. 4}
\end{figure}
%
\begin{equation}\label{eq:deterministic}
P_D^\text{\mbox{nG-AMD}}=\int_0^1P_{D|\rho}^\text{\mbox{nG-AMD}}f_\rho(\rho)d\rho
\end{equation}
%
%
%
%
Similarly, one may form the detection probability for a fluctuating target, i.e., when the distribution of $\mu_1 \sim \mathbf{x}$. 
Let
$\mathbf{x}\sim CN(0,\mathbf{R}_\mathbf{x})$. Then, the conditional detection probability of $\mbox{\mbox{nG-AMD}}$ w.r.t. $\rho$ and $\mu_1$ will have the same form as \eqref{PDrho}, i.e., $P_{D|\mu_1,\rho}^\text{\mbox{nG-AMD}}$ for a fluctuating target will have the same form as $P_{D|\rho}^\text{\mbox{nG-AMD}}$ for a deterministic target. The detection probability of \mbox{\mbox{nG-AMD}} for fluctuating target detection is
\begin{equation}\label{eq:fluctuating}
P_{D}^\text{\mbox{nG-AMD}}=\int_0^1\int_0^{+\infty}P_{D|\mu_1,\rho}^\text{\mbox{nG-AMD}}f_{\mu_1}(\mu_1)d\mu_1f_\rho(\rho)d\rho
\end{equation}
where $f_{\mu_1}(\mu_1)$ is the PDF of $\mu_1$.
Let $\mathbf{R}_0=\mathbf{A}\hat{\mathbf{R}}^{-1}\mathbf{A}$, and denote $\lambda_1\geq\lambda_2\geq...\geq\lambda_r$ the $r$ eigenvalues of $\mathbf{R}_0$, and
$\mathbf{u}_i$, for $i=1,\ldots,r$ the corresponding eigenvectors.
%
%
Furthermore, let $a_i=\lambda_i\mathbf{u}_i^H\mathbf{R}_\mathbf{x}\mathbf{u}_i$, and assume there are $m~(m<r)$ different values for $a_i$. Let $e_k$, for $k=1,\ldots,m$, denote the different values of $a_i$, i.e., $e_i\neq e_j$, for $1\leq i, j \leq m$ and $i\neq j$, with each value occurring $n_k+1$ times. Hence, $n_k\geq 0$ and $\sum_{k=1}^mn_k=r-m$.
Using \cite{LiuZSL12_60}, the PDF of $\mu_1$ may then be expressed as
\begin{align}
f_{\mu_1}(\mu_1)&=(\prod_{k=1}^mn_k!)^{-1}\sum_{k=1}^m\exp{(-\mu_1e_k^{-1})}
\notag\\
&\times
\sum_{i=1}^{n_k} \left[C_{n_k}^i\sum_{p=0}^i\text{c}(e_k, p)\mu_1^p \right]
\end{align}
for $\mu_1>0$,
where
$\text{c}(e_k,p)=C_i^pd_{n_k-i}(e_k)(r-p)^{(i-p)}e_k^{r-i-p}$,
\begin{align*}
d_{n_k-i}(e_k)&=\dfrac{\partial^{(n_k-i)}[e_k^2\prod_{i=1,i\neq k}^m({n_i}!)^{-1}(e_k-e_i)^{n_i+1}]^{-1}}{\partial e_k^{(n_k-i)}} 
%
\end{align*}
and $(n)^{(m)}= n(n-1)\cdots(n-m+1), $ for $m\geq1$, and $(n)^{(m)}=1$ for $m=0$.

\vspace{-2mm}

\section{Numerical Experiments}

This section validates the correctness of our results.
The simulation results are obtained from \eqref{eq:Lambda}, using $100/P_{\text{FA}}$ Monte Carlo simulations, whereas the theoretical results are formed by computing
$\Lambda_0$ for a desired $P_{\text{FA}}$ using \eqref{eq:pFA}, and then forming
$P_{D}$ using \eqref{eq:deterministic} or \eqref{eq:fluctuating}.
%
%
%
Let $N=6$, $r=2$, and $K=16$, with
$\mathbf{R}_{ij}=\kappa_0 0.9^{|i-j|}$, for $1\leq i,j\leq N$, where $\kappa_0$ is a scaling factor meeting the desired signal to noise ratio (SNR), here defined for a deterministic target and fluctuating target as $\text{SNR}_1=10\log_{10}({\parallel\mathbf{x}\parallel^2}/{tr(\mathbf{R})})$ and $\text{SNR}_2=10\log_{10}({tr(\mathbf{R}_\mathbf{x})}/{tr(\mathbf{R})})$, respectively.
%
%
Furthermore, the $\ell$th column of $\mathbf{A}(:,\ell)=\exp(-2j\pi f_\ell)$, for $l = 1,2,...,r$, where $f_\ell=\ell\times1.8[0:N-1]^T/N$, and the covariance matrix of the fluctuating target signal is set to be
\begin{equation}
\mathbf{R}_\mathbf{x}=
\begin{bmatrix}
1  &  0.5j\\
-0.5j  & 1
\end{bmatrix}
\end{equation}
Figures 1-2 show the curves of $P_D$ of \mbox{\mbox{nG-AMD}} for a deterministic and a fluctuating target. For the former, it is clear that the
%
%
simulation closely follows the theoretical results, proving the correctness of \eqref{eq:deterministic}.
For a fluctuating target,
the simulations only fits the theoretical results for high detection probabilities,
such as $P_D>40\%$. This deviation is due to the (approximative) numerical integration used in forming the (infinite) integral in \eqref{eq:fluctuating}; excluding the truncation error, we assume that the simulations would follow the theoretical results also for lower  detection probabilities.
%
%
Figures 3-4 show the comparisons of the theoretical performance of different detectors for deterministic target and fluctuating target detection, respectively. It can be noted that \mbox{nG-AMD} is always performing better than  ASD, whereas it has similar performance as \mbox{1S-GLRT}.
However, since the $P_\text{FA}$ of \mbox{1S-GLRT} is related with the scaling parameter $\beta$, it should be noted that \mbox{1S-GLRT} does not have the CFAR property. More precisely, the exact detection probability of \mbox{1S-GLRT} requires the exact knowledge of scaling factor $\beta$, which is not necessary for \mbox{nG-AMD}.


%

\bibliographystyle{IEEEbib}
{
\bibliography{IEEEabrv,referencesAll} }
\end{document}